\documentclass[iop,apjl,numberedappendix,appendixfloats,twocolappendix,revtex4]{emulateapj} 
\usepackage{times} 
\usepackage{latexsym,amsmath,amssymb,amsfonts} 
\usepackage{graphicx} 
\usepackage{fancyhdr}  
\usepackage{natbib} 
\usepackage{rotfloat} 
\usepackage{verbatim} 
\usepackage{rotating}   
\usepackage{url}
\usepackage{dcolumn}
\usepackage{morefloats} 
\usepackage[breaklinks,colorlinks,citecolor=blue,linkcolor=magenta]{hyperref}  
\newcommand{\hi}{\mbox{H\,{\sc i}}}

\newcommand{\zq}{$z_{\rm q}$}
\newcommand{\zg}{$z_{\rm g}$}
\newcommand{\kms}{km\,s$^{-1}$}
\newcommand{\cmsq}{cm$^{-2}$}

\newcommand{\be}{\begin{equation}}
\newcommand{\en}{\end{equation}}

\shorttitle{OH absorption at $z<0.4$} 
\shortauthors{Gupta et al.}
\begin{document}

\title{Discovery of OH absorption from a galaxy at $\lowercase{z}\sim$0.05: implications for\\ large surveys with SKA pathfinders} 

\author{N. Gupta\altaffilmark{1},  
E. Momjian\altaffilmark{2},  
R. Srianand\altaffilmark{1},  
P. Petitjean\altaffilmark{3},  
P. Noterdaeme\altaffilmark{3}, \\ 
D. Gyanchandani\altaffilmark{4},   
R. Sharma\altaffilmark{4}, and 
S. Kulkarni\altaffilmark{4}  
}    

\affil{\\ 
$^{1}$Inter-University Centre for Astronomy and Astrophysics, Post Bag 4, Ganeshkhind, Pune 411 007, India (\textcolor{blue}{ngupta@iucaa.in}) \\ 
$^{2}$ National Radio Astronomy Observatory, P.O. Box O, 1003 Lopezville Road, Socorro, NM 87801, USA \\ 
$^{3}$ UPMC-CNRS, UMR7095, Institut d'Astrophysique de Paris, F-75014 Paris, France \\ 
$^{4}$ ThoughtWorks Technologies India Private Limited, Yerawada, Pune 411 006, India \\ 
}  

\begin{abstract}

We present the first detection of OH absorption in diffuse gas at $z>0$, along with another eight stringent 
limits on OH column densities for cold atomic gas in galaxies at $0<z<0.4$. 
The absorbing gas detected towards Q0248+430 ($z_q$=1.313) originates from a tidal tail emanating from a highly star forming galaxy 
G0248+430 ($z_g$=0.0519) at an impact parameter of 15\,kpc.  
The measured column density is $N$(OH) = (6.3$\pm$0.8)$\times$10$^{13}$($\frac{T_{\rm ex}}{3.5}$)($\frac{1.0}{f_c^{\rm OH}}$)\,\cmsq, where $f_c^{\rm OH}$ and 
$T_{\rm ex}$ are the covering factor and the excitation temperature of the absorbing gas, respectively.
In our Galaxy, the column densities of OH in diffuse clouds are of the order of $N$(OH)$\sim$10$^{13-14}$\,\cmsq.
From the incidence (number per unit redshift; $n_{21}$) of \hi\ 21-cm absorbers at $0.5<z<1$ and assuming no redshift evolution, 
we estimate the incidence of OH absorbers (with log\,$N$(OH)$>$13.6) to be $n_{\rm OH}$ = $0.008^{+0.018}_{-0.008}$ at $z\sim0.1$. 
Based on this we expect to detect 10$^{+20}_{-10}$ such OH absorbers from the MeerKAT Absorption Line Survey.
Using \hi\ 21-cm and OH 1667\,MHz absorption lines detected towards Q0248+430, we estimate 
($\Delta F/F$) = (5.2$\pm$4.5)$\times 10^{-6}$, where $F \equiv g_p (\alpha^2/\mu)^{1.57}$, 
$\alpha$ $-$ the fine structure constant, $\mu$ $-$ the electron-proton mass ratio and $g_p$ $-$ the proton gyromagnetic ratio.  
This corresponds to $\Delta\alpha/\alpha$($z=0.0519$) = (1.7 $\pm$ 1.4)$\times 10^{-6}$, which is among the stringent 
constraints on the fractional variation of $\alpha$. 

\end{abstract}

\keywords{quasars: absorption lines ---  galaxies: ISM}

\section{Introduction} 
\label{sec:intro}  

The hydroxyl radical (OH) was the first interstellar molecule to be detected at radio wavelengths \citep[][]{Weinreb63}.  
Since then there have been extensive surveys of OH emission and absorption from diffuse ($N$(OH)$\sim$10$^{13-14}$\,\cmsq) and 
dense ($N$(OH)$\sim$10$^{15-16}$\,\cmsq) interstellar clouds in the Galaxy \citep[e.g.][]{Dickey81, Wannier93, Li18}, 
and along with HCO$^+$, it has emerged as one of the best indicators of H$_2$ column densities 
\citep[][]{Liszt99}.  It is most commonly observed in 18-cm 
ground state transitions which occur at rest frequencies of 1665.402 and 1667.359 (main lines), and 
1612.231 and  1720.530\,MHz (satellite lines).
The relative strengths of these lines are rarely found to be in the 
local thermodynamic equilibrium (LTE) ratios i.e.\,1612:1665:1667:1720 MHz = 1:5:9:1. They often exhibit maser emission 
in regions associated with high density and far-infrared (FIR) radiation \citep[][]{Cohen95}.   

OH Megamasers (OHMs), being a good tracer of extreme starburst activity and merger history, have also been extensively 
surveyed in luminous infrared galaxies (LIRGs), and already been detected up to $z = 0.265$ \citep[e.g.][]{Baan89, Darling02sur, Fernandez10}.  
Furthermore, the detection of main lines and the so-called {\it conjugate} behaviour of the satellite lines have been 
reported towards a handful of radio bright AGNs \citep[e.g.][]{vanLangevelde95, Darling04}. 
But OH has been very rarely searched in normal star-forming galaxies \citep[e.g.][]{Borthakur11, Zwaan15}. 
Specifically, from the literature there is only one sight line, 4C+57.23 \citep[][]{Zwaan15}, which satisfies the selection criteria 
of the study presented here. 
To date, only three {\it intervening } OH absorbers at $z>0$ are known: 
(i) J0134$-$0931 \citep[$z = 0.765$;][]{Kanekar05}; (ii) B0218+357 \citep[$z = 0.685$;][]{Kanekar03oh}; and 
(iii) PKS\,1830$-$211 \citep[$z = 0.886$;][]{Chengalur99}.  These have led to some of the most stringent ($<10^{-5}$) 
constraints on the fractional variations of fundamental constants of physics \citep[][]{Uzan11}. 
In all three cases, the absorbing gas is from a lensing galaxy and the $N$(OH)$\sim$10$^{15-16}$\,\cmsq\ i.e. similar to dense molecular clouds 
in the Galaxy. 

In this paper, we report the {\it first} survey of OH main-line absorption from cold atomic gas, as revealed by \hi\ 21-cm absorption \citep[][]{Heiles03},  
 in a sample of $z<0.4$ galaxies. Throughout this paper we use the $\Lambda$CDM 
cosmology with $\Omega_m$=0.27, $\Omega_\Lambda$=0.73 and H$_{\rm o}$=71\,\kms\,Mpc$^{-1}$. 

\begin{table*}
\caption{Details of 21-cm absorbers for OH absorption line search.}
\vspace{-0.4cm}
\begin{center}
\begin{tabular}{cccccccccccc}
\hline
\hline
Quasar             &    Galaxy           & \zq  & \zg    &  $\int\tau dv$(\hi) &  Ref.$^*$ &  Peak flux& Spectral &  Spectral        &  $\int\tau_{1667} dv$(OH)$^\Diamond$ \\  
                   &                     &      &        &                     &       &   density &   resolution              &    rms           &                          \\        
                   &                     &      &        &   (\kms)            &       &(mJy\,beam$^{-1}$)   & (\kms)    &(mJy\,beam$^{-1}$)&      (\kms)              \\ 
(1)                &      (2)            & (3)  & (4)    &    (5)              &  (6)  &     (7)    &    (8)    &       (9)        &     (10)                          \\         
\hline                                                                                                                                                                                                  
\multicolumn{10}{c}{\bf {Quasar-galaxy pairs}}   \\                     

3C232              &   NGC3067           &0.530 & 0.0049 &   0.11              &   1   &    1563             &   1.4     &    1.7           &  $<$0.006 \\ 
Q0248+430          &   G0248+430         &1.313 & 0.0519 &   0.43              &   2   &    1207             &   1.5     &    2.1           & {\bf 0.08$\pm$0.01}$^\S$ \\ 
J084957.97+510829.0& J084958.10+510826.6 &0.584 & 0.3120 &   0.95              &   3   &    200              &   0.9     &  3.9$^\dag$, 4.4$^\ddag$   & $<$0.091 \\  
J104257.58+074850.5& J104257.74+074751.3 &2.665 & 0.0332 &   0.19              &   5   &    295              &   1.5     &    1.5           &  $<$0.027 \\
J124157.54+633241.6& J124157.26+633237.6 &2.625 & 0.1430 &   2.90              &   6   &     67              &   1.6     &    1.2           &  $<$0.099 \\
J124355.78+404358.5& 124357.15+404346.5  &1.520 & 0.0169 &   2.24              &   7   &    187              &   1.4     &    1.3           &  $<$0.035 \\
J144304.53+021419.3&  Emission-lines     &1.820 & 0.3714 &   3.38              &   3   &    144              &   2.0     &    1.4           &  $<$0.059 \\
J163956.35+112758.7& J163956.38+112802.1 &0.993 & 0.0790 &   15.7              &   8   &    152              &   1.6     &    1.4$^\dag$    &  $-$      \\
\multicolumn{10}{c}{\bf {Merging galaxy pair}}   \\                     
J094221.98+062335.2&      -            & -    & 0.1230 &   49.9              &   4   &    112              &   3.3     &  1.1$^\dag$, 1.0$^\ddag$   & $<$0.070 \\  

\hline
\end{tabular}
\end{center}
$*$:  References for \hi\ absorption data $-$ 1: \citet[][]{Carilli92}; 2: This work; 3: \citet[][]{Gupta13}; 4: \citet[][]{Srianand15}; 
5: \citet[][]{Borthakur10}; 6: \citet[][]{Gupta10}; 7: \citet[][]{Gupta17gvlbi}; 8: \citet[][]{Srianand13dib}.\\ 
$\dag$, $\ddag$: rms in the subband covering 1665 and 1667\,MHz lines, respectively; $\Diamond$: integrated OH optical depth of 1667\,MHz line or, in case of 
non-detections, 3$\sigma$ upper limit for a spectral resolution of 2\,\kms.   ; $\S$: the 3$\sigma$ optical depth limit is 0.009\,\kms.\\

\label{ohsamp}
\end{table*}

\section{Sample, observations and data analysis}      
\label{sec:obs}   

We used the Giant Metrewave Radio Telescope (GMRT), the Karl G. Jansky Very Large Array (VLA) and the Westerbork Synthesis Radio Telescope (WSRT)  
to search for OH 18-cm main lines in 9 \hi\ 21-cm absorbers.  In 8 cases, the intervening 21-cm absorption originates from cold atomic gas associated with galaxies 
at $z<0.4$.  The names of background quasar and foreground galaxy which we refer to as quasar-galaxy pairs (QGPs), their redshifts and 
\hi\ 21-cm optical depths are provided in Table~\ref{ohsamp}. 
This sample essentially represents all the $z<0.4$ \hi\ 21-cm absorption in QGPs that were known in 2013 either from the literature or our own survey 
\citep[see][for the latest]{Dutta17}, and considering the radio frequency interference (RFI) environment, could be observed for OH 
with the above-mentioned telescopes. 
In the case of J1443+0214, the absorption is associated with a low surface brightness galaxy that, unlike other QGPs is identified only via narrow 
optical emission lines detected on top of the QSO spectrum.
We also included the associated 21-cm absorption from the merging pair J0942+0623.  This is one of the strongest \hi\ 21-cm absorbers 
\citep[$N$(\hi)$\sim$10$^{22}$\,cm$^{-2}$;][]{Srianand15}, and hence, a promising candidate for molecular line search.

Five quasars, namely 3C232, Q0248+430, J1042+0748, J1241+6332 and J1243+4043, were observed with the WSRT in Maxi-short configuration in 
December 2010.  
A baseband bandwidth of 10 MHz split into 2048 frequency channels was used.  The on-source time was $\sim$10\,hrs per QGP.  
J1443+4043 was observed with the GMRT for $\sim$6\,hrs  on 2012, June, 2 using a bandwidth of $\sim$4\,MHz split into 512 channels. 
The remaining three quasars were observed with the VLA in A or A$\rightarrow$D configurations over October 2012 - January 2013.  
The WIDAR correlator was set-up to split 16 subbands into 128 frequency channels. The subband bandwidths of 0.5 MHz, 2 MHz and 1 MHz were used for 
J0849+5108, J0942+0623 and J1639+1127, respectively.  The total on-source time was 3-8\,hrs per QGP. 
For GMRT and WSRT observations, same baseband covered both the main lines, whereas for VLA observations these were placed in two separate subbands.

The VLA and GMRT data were reduced using the Automated Radio Telescope Imaging Pipeline \citep[{\tt ARTIP};][]{artip18}.   
The WSRT data were calibrated using {\tt AIPS}.  The continuum and line imaging including self-calibration were performed using  
{\tt ARTIP}. For Q0248+430, the continuum-subtracted spectral line cube was deconvolved using CLEAN down to twice the single channel noise using 
image masks.  The OHM emission and OH 1667\,MHz absorption detected from this QGP are spatially separated by 15$^{\prime\prime}$, and 
by $\sim$116\,km\,s$^{-1}$ in velocity space.
The CLEANing mask included pixels with OHM emission, and the region over which the radio continuum from the quasar is detected.  
We also examined the channel maps to ensure that none of the detected signal is due to residual deconvolution errors or 
continuum subtraction.

The stokes $I$ peak flux densities of quasars are provided in Table~\ref{ohsamp}. The spatial resolution of GMRT and VLA images is 
2-3$^{\prime\prime}$. For J1042+0748 and the remaining WSRT maps the spatial resolution is $190^{\prime\prime}\times10^{\prime\prime}$ 
and $\sim15^{\prime\prime}$, respectively.      
The spectral resolution and the corresponding root-mean-square (RMS) values are in the range of 0.9\,-\,3.3\,\kms\ and 1.2\,-\,4.4\,mJy\,beam$^{-1}$, 
respectively (see columns 8-9 of Table~\ref{ohsamp}).  
For J1639+1127, the VLA subband covering 1667\,MHz line was affected by RFI.  The 3$\sigma$ optical depth limit for the 1665\,MHz line is 0.050\,\kms.  

 
\begin{figure} 
\centerline{\vbox{
\centerline{\hbox{ 
\includegraphics[width=0.50\textwidth,angle=0]{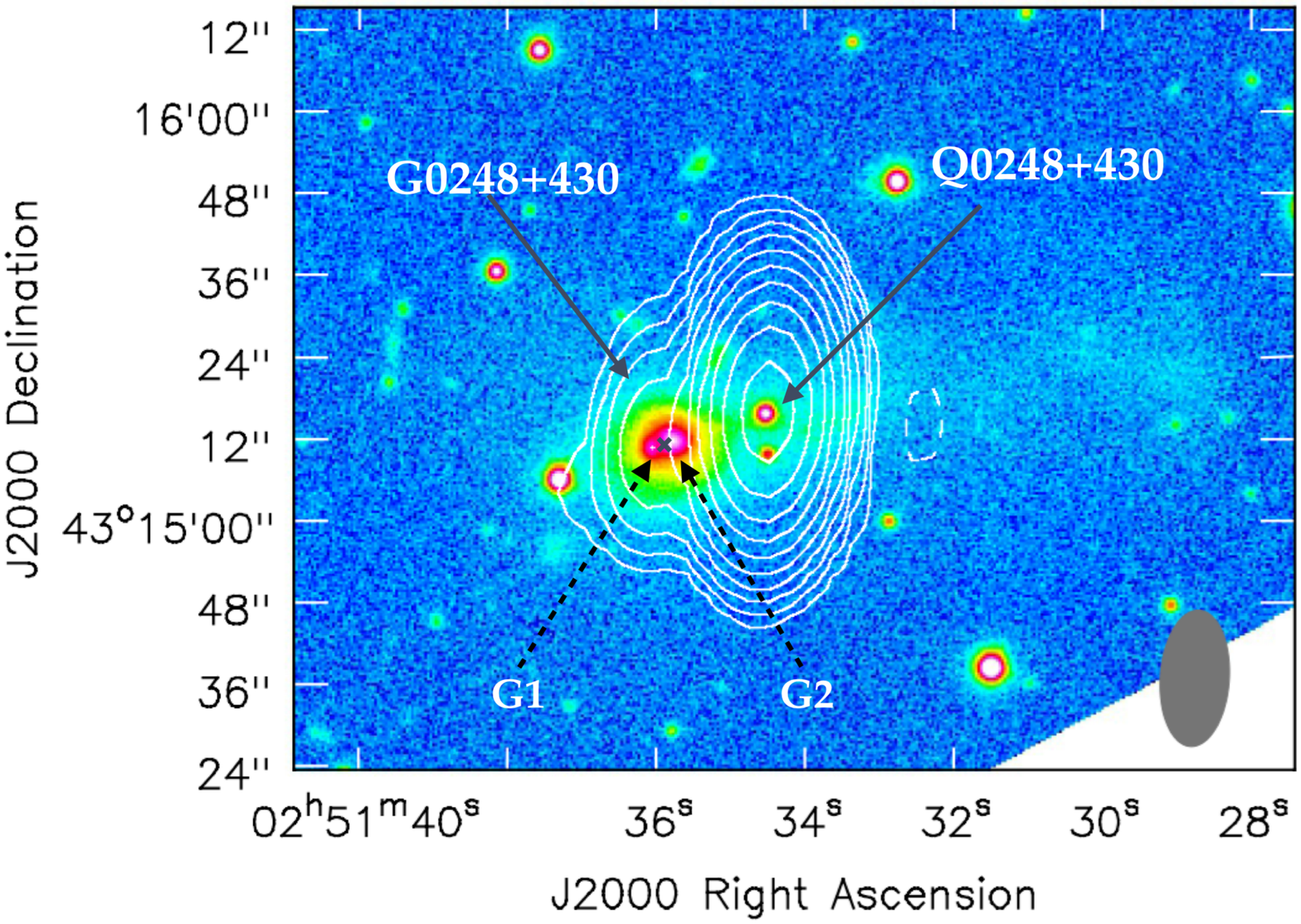}  
}} 
}}  
\vskip+0.2cm  
\caption{The WSRT radio continuum ($\sim$1590\,MHz) contours overlaid on the SDSS r-band image of the QGP 0248+430.  The contour levels are 2$\times$2$^n$\,mJy\,beam$^{-1}$ 
(where n=-1,0,1,2,3,...).  The restoring beam of 20.1$^{\prime\prime}$$\times$10.2$^{\prime\prime}$ with position angle -3.8$^\circ$ is 
also shown. The integrated flux densities of radio components associated with the quasar and the galaxy are 
1216 and 24\,mJy, respectively.  The location of OHM emission ($\alpha$(J2000)=02:51:35.945, $\delta$(J2000)=43:15:11.00) is marked with a $\times$.  
} 
\label{fig:gal}   
\end{figure} 

\begin{figure*} 
\centerline{\vbox{
\centerline{\hbox{ 
\hspace*{-1cm}\includegraphics[trim = {0cm 0cm 0cm 0cm}, clip, width=0.7\textwidth,angle=-00]{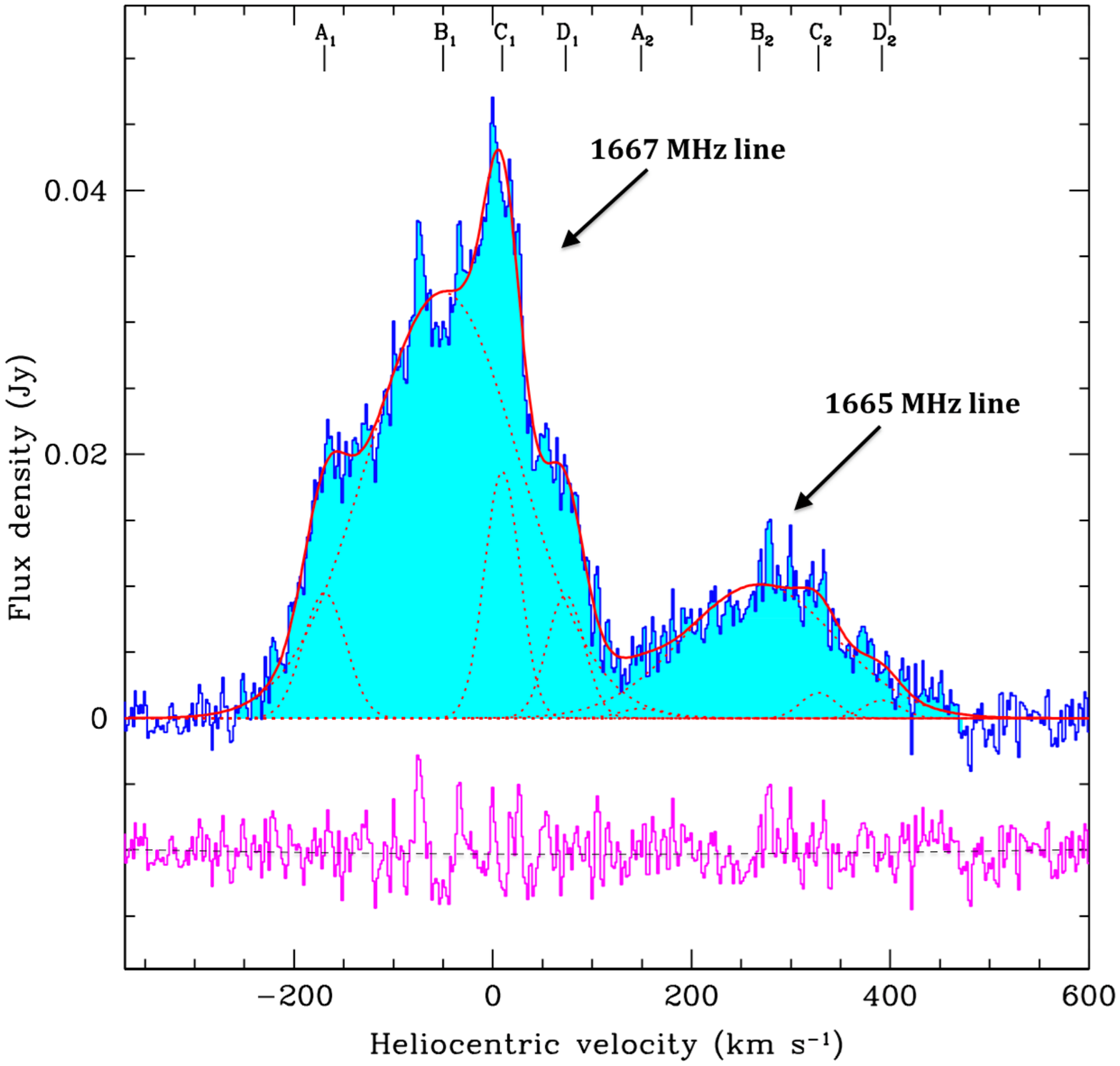}  
\hspace*{-2cm}\includegraphics[trim = {0cm 5cm 0cm 0cm}, clip, width=0.53\textwidth,angle=-00]{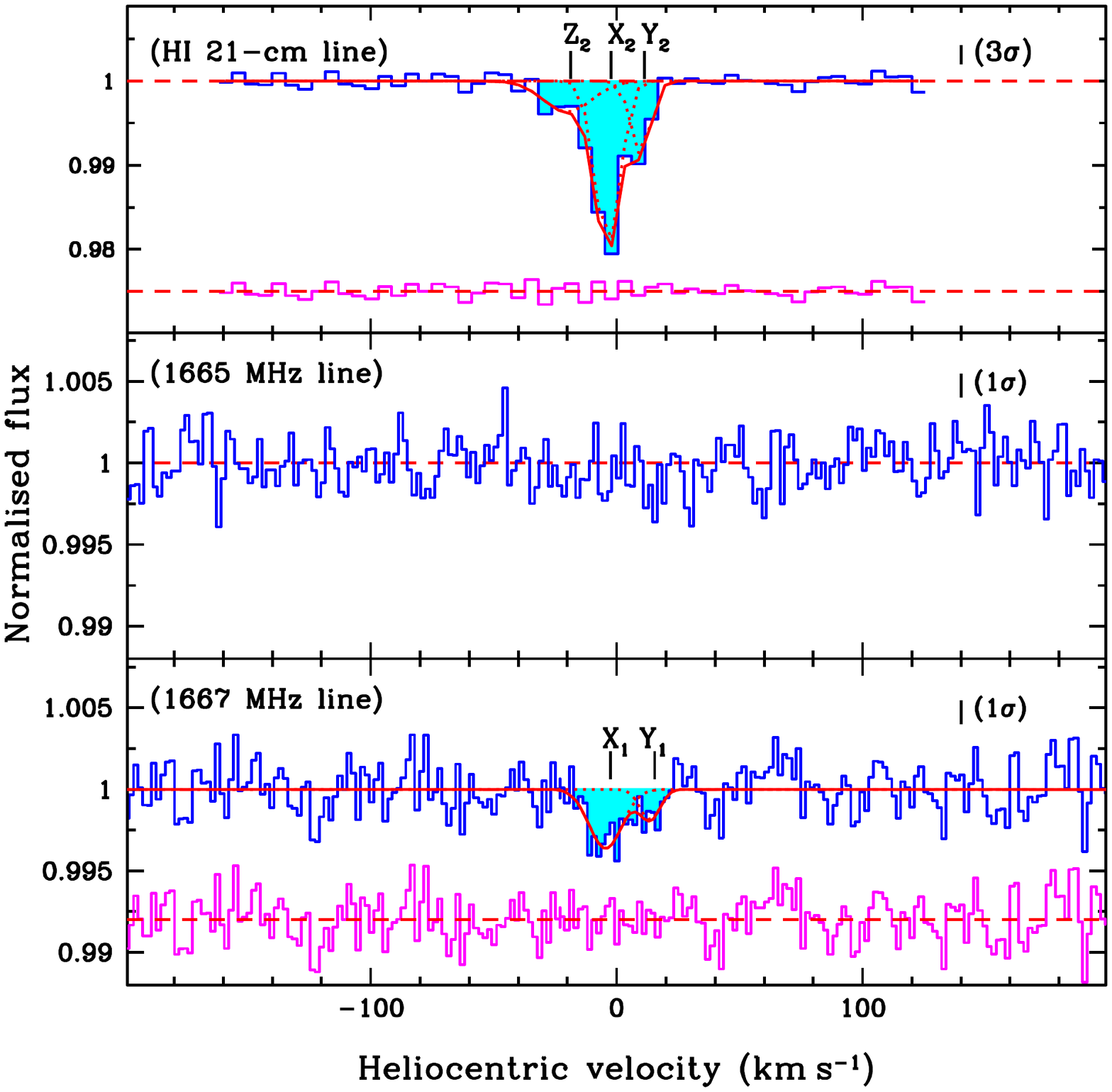}  
}} 
}}  
\vskip-0.2cm  
\caption{
{\it Left:}
OHM detected in G0248+430.  
The zero of the velocity scale ($z = 0.05192$) is centered at the peak of the 1667\,MHz line.   
{\it Right:} \hi\ 21-cm and OH main line absorption spectra towards Q0248+430.  The zero of the velocity scale ($z = 0.05151$) is centered at the peak of
\hi\ 21-cm absorption line.
In both the panels, individual Gaussian components (see Table~\ref{ohmgauss}), the resulting fits to the stokes-$I$ spectrum are plotted as dotted and 
continuous lines, respectively. The residuals, on an offset arbitrarily shifted for clarity, are also shown.
} 
\label{fig:ohdet}   
\end{figure*} 

\section{OH detection at $\lowercase{z}=0.05$: QGP\,0248+430}    
\label{sec:OH}  

The radio-optical overlay of this QGP is shown in Fig.~\ref{fig:gal}. 
The QSO Q0248+430 ($z_q=1.313$) is at an angular separation of 15$^{\prime\prime}$ ($\sim$15\,kpc at $z_g\sim$0.05) from the foreground galaxy G0248+430.  
The latter is actually a pair of merging 
spiral ($z = 0.0512 \pm 0.0010$) and  elliptical ($z = 0.0507 \pm 0.0010$) galaxies separated by 3.5$^{\prime\prime}$ 
\citep[3.5\,kpc;][]{Kollatschny91}.  They are 
labelled as G1 and G2, respectively, in Fig.~\ref{fig:gal} and  
constitute a rare system where optical emission lines from both the nuclei indicate non-thermal activity   \citep[][]{Borgeest91}.  
The tidal tail emanating from G0248+430 has bluer colors compared to the galaxy and extends across the line of sight to Q0248+430 
\citep[see Fig.~\ref{fig:gal};][]{Borgeest91}.  It also shows stellar absorption lines confirming the presence of star formation 
activity in the tail. 

We detect OHM emission coincident with the central region of the galaxy. The OH absorption towards the background quasar  is detected from 
the gas associated with the tidal tail.   

\subsection{OHM emission in G0248+430}    
\label{sec:ohem}  

The optical imaging, spectroscopy and strong FIR emission suggest that the system has undergone a very recent and strong starburst.  
We estimate the IR luminosity of the galaxy using the flux densities from the IRAS Faint Source Catalog \citep[][]{Moshir90}, 
$f[12\mu m, 25\mu m, 100\mu m, 600\mu m]$ = [$<$6.54$\times$10$^{-2}$, 0.19 $\pm$ 0.01, 4.02 $\pm$ 0.28, 6.92 $\pm$ 0.42]\,Jy and
the following equation from \citet{Sanders96}, 
\begin{equation}
\begin{split}
L_{\rm IR} (L_ {\odot})=5.67 \times 10^{5}~D_{\rm L}^{2}~(13.48f_{12} + 5.16f_{25} + \\ 2.58f_{60} + f_{100}),
\end{split}
\end{equation}
where $D_{\rm L}$ is the luminosity distance in Mpc, to be  $L_{\rm IR} < 5.81 \times 10^{11}\,L_ {\odot}$. 

Highly luminous IR galaxies such as G0248+430 are often associated with OHMs.
\citet[][]{Kazes89} report the detection of OHM from this galaxy but do not provide any detail.     
We present OHM emission from the galaxy in  Fig.~\ref{fig:ohdet}. (see also `$\times$' in Fig.~\ref{fig:gal}). 
Both the main lines are detected.  To separate the contributions of two hyperfine lines, 
we model these using multiple Gaussian components. 
We note that Gaussian component fitting provides a convenient measure of source/spectral structure even if they do not necessarily represent discrete physical structures.
 Under the assumption that both the 
lines originate from the same gas, the centres and widths of the components for 1665\,MHz line are tied to those of 1667\,MHz line.  
The overall structure of 1667\,MHz line is reasonably modelled by a four-component fit. As discussed below, we believe 
that the remaining structure in the residuals is an artefact of the limited spatial resolution 
($\sim$20$\times$10\,kpc$^2$) of the WSRT image.  Therefore, we do not attempt to improve it further by adding more components.   
Noticeably, the components A$_2$, C$_2$ and D$_2$ are barely detected and contribute only $\sim$7\% to the total integrated flux density of 
the 1665\,MHz line (see Table~\ref{ohmgauss}). 
The integrated flux densities of the 1665 and 1667\,MHz lines are 2.1 and 8.0\,Jy\,\kms, respectively.  
These correspond to a total OH luminosity of 860\,$L_\odot$. This is about a factor 6 higher than the luminosity 
expected from the $L_{\rm FIR} - L_{\rm OH}$ correlation \citep[cf. equation 4 of][]{Darling02sur} but not unusual 
or significant considering the statistical scatter in the relationship. 
Here, the FIR luminosity is estimated using the following equation from  \citet[][]{Sanders96}
\begin{equation}
L_{\rm FIR} (L_ {\odot}) = 3.96 \times 10^{5}~D_{\rm L}^{2}~(2.58f_{60}+f_{100}),
\end{equation}
to be $L_{\rm FIR} = 3.50 \times 10^{11}\,L_ {\odot}$.

The hyperfine ratio of the observed 1667 to 1665\,MHz line integrated flux densities is 3.8.  The ratios of the flux densities of individual Gaussian 
components fitted to these lines are also far from the value of 1.8 expected for LTE. 
In general, Very Long Baseline Interferometry (VLBI) observations often resolve OHMs into multiple components 
\citep[e.g.][]{Momjian06}. 
The detailed modeling of these suggests that the observed differences between 1667 to 1665\,MHz line ratios can be explained by the 
exponential amplification of the background radiation by unsaturated maser clouds overlapping in space and velocity 
\citep[e.g.][]{Parra05, Lockett08, Willett11}.

\begin{table}
\caption{Multiple Gaussian fits to the OH and \hi\ lines.}
\vspace{-0.4cm}
\begin{center}
\begin{tabular}{ccccccccc}
\hline
  \multicolumn{4}{c}{$\Longleftarrow$ ~~~~~~~~~OHM 1667\,MHz~~~~~~ $\Longrightarrow$}   &                     
  \multicolumn{3}{c}{$\Longleftarrow$  OHM 1665\,MHz  $\Longrightarrow$}                        \\
{\large \strut}     Id.    &    Centre     &  $\sigma$            &   Peak         &   Id.                &    Centre   &   Peak      \\
                           &   (\kms)      &   (\kms)             &   (mJy)        &                      &    (\kms)   &  (mJy)      \\
                  (1)      &   (2)         &    (3)               &   (4)          &   (5)                &    (6)      &   (7)       \\
\hline
                A$_1$      & -169 $\pm$ 1    & 22 $\pm$ 1   & 9.5 $\pm$ 0.5   &  A$_2$   &  149  & 0.7 $\pm$ 0.5          \\
                B$_1$      &  -50 $\pm$ 2    & 77 $\pm$ 2   &32.3 $\pm$ 0.3   &  B$_2$   &  268  &10.1 $\pm$ 0.2         \\
                C$_1$      &   10 $\pm$ 1    & 18 $\pm$ 1   &18.7 $\pm$ 0.8   &  C$_2$   &  328  & 1.9 $\pm$ 0.5         \\
                D$_1$      &   73 $\pm$ 1    & 19 $\pm$ 2   & 9.3 $\pm$ 0.9   &  D$_2$   &  392  & 1.4 $\pm$ 0.5         \\
\hline

  \multicolumn{4}{c}{$\Longleftarrow$  OH 1667\,MHz absorption $\Longrightarrow$}   &                     
  \multicolumn{3}{c}{$\Longleftarrow$  \hi\ 21-cm absorption $\Longrightarrow$}                        \\
{\large \strut}     Id.    &    Centre     &  $\sigma$            &   $\tau_p$     &   Id.                &    Centre   &  $\tau_p$      \\
                           &   (\kms)      &   (\kms)             &   ($10^{-3}$)  &                      &    (\kms)   &  ($10^{-3}$)      \\
\hline
                X$_1$      &   -4$\pm$2      & 7$\pm$2      & 4  $\pm$ 1 &  X$_2$ &  -4$\pm$1   & 20 $\pm$ 2             \\
                Y$_1$      &   14$\pm$3      & 4$\pm$2      & 2  $\pm$ 1 &  Y$_2$ &  10$\pm$1   &  9 $\pm$ 1            \\
                  -        &       -         &    -         &      -     &  Z$_2$ & -20$\pm$4   &  3 $\pm$ 1            \\
\hline
\end{tabular}
\label{ohmgauss}
\end{center}
$\tau_p$: peak optical depth; the fits to OH absorption are used only for the analysis in Section~\ref{sec:const}.
\end{table}

\subsection{OH absorption from G0248+430}    
\label{sec:ohabs}  

The presence of cold atomic gas and metals in the tidal tail emanating from G0248+430 have been inferred from \hi\ 21-cm, Na~{\sc i} and Ca~{\sc ii} 
absorption detections towards Q0248+430 \citep[][]{Womble90, Hwang04}. The ratio $N$(Ca~{\sc ii})/$N$(Na~{\sc i}) is 
similar to the values observed in the Galactic disk.
Here we report the detection of OH absorption towards Q0248+430 implying the presence of molecular gas in the tidal tail.
The stokes-$I$ OH absorption spectra are shown in  Fig.~\ref{fig:ohdet}.  
Only the  1667\,MHz line is detected.  The total optical depth obtained by integrating over the absorption profile
is $\int\tau_{1667} dv$(OH) = 0.08$\pm$0.01\,\kms. 
The absorption is also consistently reproduced in individual XX and YY spectra. 
We integrate 1665\,MHz spectrum over -20 to +20\,\kms\ and obtain 
$\int\tau_{1665} dv$(OH) = 0.04$\pm$0.01\,\kms.  We consider this to be a non-detection with an upper limit on 
the integrated optical depth, $\int\tau$dv$<$0.04\,\kms. This will be consistent with it being subthermal or in LTE.
For an optically thin cloud, the integrated OH optical depth of the 1667\,MHz line is related to the OH column density $N$(OH) through,
\begin{equation}
N{({\rm OH})}=2.24\times10^{14}~{T_{\rm ex}\over f_{\rm c}^{\rm OH}}\int~\tau_{1667}(v)~{\rm d}v~{\rm cm^{-2}}, 
\label{eq1}
\end{equation}
where $T_{\rm ex}$ is the excitation temperature in Kelvins, $\tau_{1667}$($v$) is the optical depth of the 1667\,MHz line at velocity $v$, 
and $f_c^{\rm OH}$ is the covering factor \citep[e.g.][]{Liszt96}. 
For Q0248+430, adopting $f_c^{\rm OH}$ = 1, $T_{\rm ex}$ = 3.5\,K which is the peak of the log-normal function fitted to the 
$T_{ex}$ distribution of OH absorbers observed in the Galaxy \citep[][]{Li18} and 
$\int\tau_{1667} dv$(OH) = 0.08$\pm$0.01\,\kms\ from Table~\ref{ohsamp}, 
we get $N$(OH) = (6.3$\pm$0.8)$\times$10$^{13}$($\frac{T_{\rm ex}}{3.5}$)($\frac{1.0}{f_c^{\rm OH}}$)\,\cmsq.  This is similar to the $N$(OH)$\sim$10$^{13-14}$\,\cmsq\ 
observed in diffuse clouds in the Galaxy, 
but 15$-$550 times lower than the column densities of three previously known intervening OH absorbers from gravitational lenses (cf. Section~\ref{sec:ndet}).  

We reprocessed the archival VLA data used for \hi\ 21-cm absorption analysis in \citet[][]{Hwang04}. We measure the peak flux density to be 
944\,mJy\,beam$^{-1}$ and the total integrated 21-cm optical depth measured from the spectrum, $\int\tau_{21}dv$ = 0.43$\pm$0.02\,kms (see Fig.~\ref{fig:ohdet}). 
For an optically thin cloud the $\int\tau_{21}dv$ is related to the neutral 
hydrogen column density $N$(H~{\sc i}), spin temperature $T_{\rm s}$, and covering factor $f_c^{\tiny \hi}$ through,
\begin{equation}
N{(\hi)}=1.823\times10^{18}~{T_{\rm s}\over f_{\rm c}^{\tiny \hi}}\int~\tau(v)~{\rm d}v~{\rm cm^{-2}}.
\label{eq1}
\end{equation}
For $f_c^{\tiny \hi} =$1, as discussed below and adopting $T_{\rm s}=$70\,K, 
which is the median column density weighted $T_{\rm s}$ for the cold neutral medium (CNM) in our 
Galaxy \citep[][]{Heiles03}, we get $N$(\hi) = (5.5$\pm$0.3)$\times10^{19}$($T_{\rm s}$/70)(1.0/$f_c^{\tiny \hi}$)\,cm$^{-2}$. 
The [OH]/[\hi] abundance ratio for Q0248+430 is $10^{-6}$, which although not unusual  
is about an order of magnitude higher than the typical ratio ($\sim$10$^{-7}$) observed in the Galaxy \citep[][]{Li18}.  
A much larger $T_{\rm s}$ ($\sim$1000\,K), as is more commonly seen in $z>2$ \hi\ absorbers \citep[][]{Srianand12dla, Kanekar14}, 
and/or $f_c^{\rm OH}<1$ would give [OH]/[\hi] more in accord with Galactic observations.

In the VLBI image at 2.3\,GHz, Q0248+430 is resolved into multiple components with an overall extent of 26\,milliarcsecond 
\citep[27\,pc at $z_g=0.05$;][]{Fey00}.  
VLBI spectroscopic observations of low-$z$ \hi\ 21-cm absorbers show that for diffuse ISM the extent of CNM gas is 
$>$20\,pc \citep[][]{Keeney05, Gupta17gvlbi}.  
Therefore, it is quite likely that  the \hi\ absorber in front of Q0248+430 fully covers the radio emission i.e. $f_c^{\tiny \hi} =1$. 
However, sizes of diffuse H$_2$ components associated with damped Lyman-$\alpha$ systems (DLAs) have been inferred to be $<15$\,pc 
\citep[][]{Srianand12dla, Noterdaeme17}, implying that probably $f_c^{\rm OH}<1$.

\subsection{Variation of fundamental constants}    
\label{sec:const}  

As OH and \hi\ absorption line frequencies depend differently on $\alpha$ $-$ the fine structure constant, $\mu$ $-$ 
the electron-proton mass ratio and $g_p$ $-$ the proton gyromagnetic ratio, relative shifts between the observed 
frequencies of these lines can be used to constrain the variations of these fundamental constants of physics. 
But this crucially requires that both the absorption lines originate from the same gas.  
For Q0248+430, the \hi\ 21-cm absorption is broader compared to the OH absorption. Specifically, the 90\% of the total 
OH and \hi\ optical depths are contained within 30$\pm$1 and 43$\pm$5\,\kms, respectively. 
The 1667\,MHz line clearly shows two absorption components.  We use two Gaussian components, X$_1$ and Y$_1$, to model 
it and determine the frequencies of the two peaks (Table~\ref{ohmgauss}).   
The 21-cm line has much higher signal-to-noise ratio (SNR) but the spectral resolution is coarser by a factor of four.  
Besides absorption corresponding to X$_1$ and Y$_1$, the 21-cm line also has an additional absorption component at -20\,\kms.  
Therefore, we model it using three components X$_2$, Y$_2$ and Z$_2$, with the widths of first two components fixed to X$_1$ 
and Y$_1$, respectively (Table~\ref{ohmgauss}).  We note that the broad 21-cm absorption ($\sigma=10\pm3$) corresponding to Z$_2$ is 
reported here for the first time \citep[cf. Fig.~7 of][]{Hwang04}.  

We next compare redshifted frequencies of \hi\ and OH absorption components, X$_i$ and Y$_i$, for $i$=1 and 2, to 
constrain $\Delta F/F$  = ($z_{\rm OH} - z_{\small \hi}$)/(1 + $z_{\rm OH}$), where $F \equiv g_p (\alpha^2/\mu)^{1.57}$ \citep[][]{Uzan11}. 
The centres of components fitted to the OH and \hi\ are: 
$z_{\rm X_1}$ = 0.051498 $\pm$ 0.000007, 
$z_{\rm Y_1}$ = 0.051561 $\pm$ 0.000007,
$z_{\rm X_2}$ = 0.051497 $\pm$ 0.000002, and
$z_{\rm Y_2}$ = 0.051551 $\pm$ 0.000002. 
This yields for the two components, ($\Delta F/F$)$_X$ = (0.95 $\pm$ 6.3)$\times 10^{-6}$ and ($\Delta F/F$)$_Y$ = (9.5 $\pm$ 6.3)$\times 10^{-6}$.  
The weighted average of these provides:   ($\Delta F/F$) = (5.2 $\pm$ 4.5)$\times 10^{-6}$.  
Taking the case of $\alpha$ as it has strongest dependence on $F$, and assuming that $\mu$ and $g_p$ are constant, we get 
$\Delta\alpha/\alpha$($z=0.052$) = 
(1.7 $\pm$ 1.4)$\times 10^{-6}$.  This is among the stringent constraints on the variation of $\alpha$ \citep[][]{Rahmani12, Kanekar18}.  

The constraints on $\mu$ and $g_p$ will be weaker but more importantly, here,  
we assumed that the components X and Y for OH and \hi\ absorption are tracing the gas with same physical conditions and internal motions 
within the cloud.  The same is implicitly assumed for the component Z which is only detected in the higher SNR \hi\ spectrum.  This and the 
uncertainty due to $f_c^{OH} \neq f_c^{\small \hi}$ caused either by the different sizes of \hi\ and OH clouds or the proper motion of the 
radio source components between the epochs of \hi\ and OH observations are the major unaccounted sources of errors in our analysis. 
More sensitive near-simultaneous observations of this absorber especially using the Very Long Baseline Interferometry (VLBI) are planned to address these uncertainties. 
  
\section{Implications of OH non-detections}    
\label{sec:ndet}  
\citet[][]{Li18} recently published observations of OH main lines from the Galaxy towards 44 extragalactic continuum sources.
The \hi\ and OH column densities derived by them from the Gaussian component-by-component analysis along with the measurements from our 
survey are presented in Fig.~\ref{fig:ohplot}. 
The {\it top} panel of Fig.~\ref{fig:ohplot}, provides detection rate $R_{Gal}$ for different limiting values of OH and \hi\ column 
densities, $N$({\rm OH})$_{lim}$ and $N$(\hi)$_{lim}$, respectively. For given values of these, $R_{Gal}$ is estimated 
by determining the number of OH detections with OH and \hi\ column densities larger than the limiting values, and dividing it by total number 
of sight lines with $N$(\hi)$\ge$$N$(\hi)$_{lim}$ but considering only those that are sensitive to detect OH down to $N$({\rm OH})$_{lim}$.   
It is apparent from the figure that 
(i) $R_{Gal}$ is smaller for larger $N$({\rm OH})$_{lim}$ i.e. absorbers with larger $N$(OH) are rarer, in particular none similar to 
known absorbers from Gravitational lenses are detected, and 
(ii) $R_{Gal}$ does not depend on $N$({\hi)$_{lim}$.
In general, the trends in $R_{Gal}$ with respect to $N$({\rm OH})$_{lim}$ or $N$(\hi)$_{lim}$ can shed light on the nature of OH absorbers and 
the conditions in which they are likely to be detected. Due to large statistical errors we are unable to draw any substantial conclusions. 
At this point, these Galactic measurements provide the minimal context in which to view the detections/non-detections from our survey.

\begin{figure} 
\centerline{\vbox{
\centerline{\hbox{ 
\includegraphics[trim = {0 5cm 0 2cm}, clip, width=0.48\textwidth,angle=0]{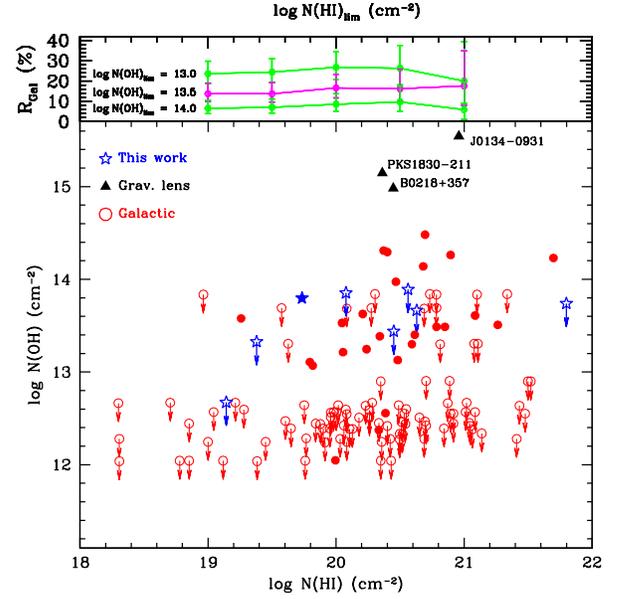}  
}} 
}}  
\vskip+0.2cm  
\caption{$N$(OH) vs $N$(\hi) for \hi\ 21-cm absorbers.  $\star$, $\triangle$ and $\circ$ represent measurements for known \hi\ 21-cm absorbers from this work, 
three known extragalactic OH absorbers from gravitational lenses and Galactic measurements from \citet[][]{Li18}, respectively. Filled symbols are for 
OH detections.  The detection rates ($100\times R_{Gal}$) for only Galactic measurements are plotted in top panel.  
All the $N$(OH) upper limits have been estimated adopting $f_c$=1, $T_{\rm ex}$ = 3.5\,K and line FWHM = 2\,\kms.  To estimate $N$(\hi) for 
extragalactic \hi\ 21-cm absorbers, $T_{\rm s}$ = 70\,K is assumed.
} 
\vspace{+0.4cm}
\label{fig:ohplot}   
\end{figure} 

The $N$(OH) upper limits from our sample are in the range: 0.4\,-\,7.7$\times$10$^{13}$\,\cmsq.  These are sensitive to detect OH 
at higher end of column densities observed in the diffuse ISM (refer to Galactic measurements in Fig.~\ref{fig:ohplot}). 
We also note that the highest $N$(\hi) absorber in our sample is a non-detection in OH. The associated 
AGN (J0942+0632) in this case is resolved into multiple components extending over 89\,pc \citep[][]{Srianand15}.  The OH non-detection could be due 
to $f_c^{\rm OH} \ll$1, or that there is no molecular gas along the sight line.        

Next we use the method used to estimate $R_{Gal}$ to obtain the detection rate, $R_{sur}$,  for our OH survey.
For this we adopt log\,$N$({\rm OH})$_{lim}$(\cmsq) = 13.7 and log\,$N$(\hi)$\ge$$N$(\hi)$_{lim}$(\cmsq) = 19.0.
We estimate $R_{sur}$ = 1/4 = 25$^{+58}_{-21}$\%.  Note that this does not include the measurements for three gravitational 
lens systems which have column densities in the range of dense molecular gas. 
From \citet[][]{Gupta12}, $n_{21}$ for integrated 21-cm optical depths of 0.1-0.5\,\kms are available. We note that for 70K, 0.1\,\kms\ corresponds 
to log\,$N$(\hi)(\cmsq) = 19.1.
Adopting, $n_{21}(0.5<z<1)$ = 0.03$^{+0.03}_{-0.02}$ from \citet[][]{Gupta12} for log\,$N$(\hi)$\ge$$N$(\hi)$_{lim}$(\cmsq) = 19.0 and assuming no redshift evolution in $n_{21}$, 
we get the number per unit redshift range of OH absorbers, $n_{\rm OH}$ = $R\times n_{21}$ = $0.008^{+0.018}_{-0.008}$. 

Although derived for somewhat arbitrarily chosen limiting values of $N$(\hi) and $N$(OH)  to include most of the measurements from our survey, 
the adopted limits are relevant for the upcoming MeerKAT Absorption Line Survey \citep[MALS;][]{Gupta17mals} which will have the sensitivity to detect 
cold atomic and molecular gas with $N$(\hi)$>10^{19}$\,\cmsq and $N$(\hi)$>10^{14}$\,\cmsq. 
Based on the derived $n_{\rm OH}$ and the total MALS redshift path of $\sim$1000 towards on-axis (primary) strong radio sources, we expect to 
detect 10$^{+20}_{-10}$ OH absorbers from diffuse ISM of external galaxies.

\section{Concluding remarks}    
\label{sec:conc}  

We have used GMRT, VLA and WSRT to perform the first survey of OH absorption from cold atomic gas in galaxies. 
The survey has led to first detection of OH from diffuse molecular gas ($N$(OH)$\sim$10$^{13-14}$\,\cmsq) at $z>0$.  
The absorber is the first one to enable further detailed studies through VLBI spectroscopy to improve our understanding 
of the physical extent of both cold atomic and molecular gas, and directly address the systematics affecting the constraints 
on fundamental constants of physics through radio absorption lines.
The three previously known intervening OH absorbers are at higher $z$ and due to unavailability of suitable low-frequency receivers 
can not be observed through VLBI spectroscopy. 
A substantial number of OH absorbers may be detected from large surveys with Square Kilometer Array (SKA) pathfinders 
allowing these to be used as an effective tool to probe complex gas physics and variations of fundamental constants.
The majority of radio absorption line surveys till now have been based on sight lines from optical spectroscopic surveys which are 
biased against dust. They are indeed tracing diffuse ISM. This will change with upcoming blind radio absorption line surveys which 
will trace both diffuse and dense ISM. The results from the survey presented here will still be applicable to the part of the 
survey(s) tracing diffuse ISM.

\acknowledgments

We thank the referee for useful comments.
NG, PN, PPJ and RS acknowledge support from the Indo-French Centre for the Promotion of Advanced Research 
under Project {\tt 5504-B}.
GMRT is run by the National Centre for Radio Astrophysics of the Tata Institute of Fundamental Research.
The VLA is run by the National Radio Astronomy Observatory which is a facility of the National Science Foundation 
operated under cooperative agreement by Associated Universities, Inc. 
WSRT is operated by the ASTRON (Netherlands Institute for Radio Astronomy) with 
support from the Netherlands Foundation for Scientific Research (NWO).



\begin{thebibliography}{}
\expandafter\ifx\csname natexlab\endcsname\relax\def\natexlab#1{#1}\fi
\providecommand{\url}[1]{\href{#1}{#1}}

\bibitem[{{Baan}(1989)}]{Baan89}
{Baan}, W.~A. 1989, \apj, 338, 804

\bibitem[{{Borgeest} {et~al.}(1991){Borgeest}, {Schramm}, {Dietrich},
  {Kollatschny}, \& {Hopp}}]{Borgeest91}
{Borgeest}, U., {Schramm}, K.-J., {Dietrich}, M., {Kollatschny}, W., \& {Hopp},
  U. 1991, \aap, 243, 93

\bibitem[{{Borthakur} {et~al.}(2011){Borthakur}, {Tripp}, {Yun}, {Bowen},
  {Meiring}, {York}, \& {Momjian}}]{Borthakur11}
{Borthakur}, S., {Tripp}, T.~M., {Yun}, M.~S., {et~al.} 2011, \apj, 727, 52

\bibitem[{{Borthakur} {et~al.}(2010){Borthakur}, {Tripp}, {Yun}, {Momjian},
  {Meiring}, {Bowen}, \& {York}}]{Borthakur10}
---. 2010, \apj, 713, 131

\bibitem[{{Carilli} \& {van Gorkom}(1992)}]{Carilli92}
{Carilli}, C.~L., \& {van Gorkom}, J.~H. 1992, \apj, 399, 373

\bibitem[{{Chengalur} {et~al.}(1999){Chengalur}, {de Bruyn}, \&
  {Narasimha}}]{Chengalur99}
{Chengalur}, J.~N., {de Bruyn}, A.~G., \& {Narasimha}, D. 1999, \aap, 343, L79

\bibitem[{{Chengalur} \& {Kanekar}(2003)}]{Kanekar03oh}
{Chengalur}, J.~N., \& {Kanekar}, N. 2003, Physical Review Letters, 91, 241302

\bibitem[{{Cohen}(1995)}]{Cohen95}
{Cohen}, R.~J. 1995, \apss, 224, 55

\bibitem[{{Darling}(2004)}]{Darling04}
{Darling}, J. 2004, \apj, 612, 58

\bibitem[{{Darling} \& {Giovanelli}(2002)}]{Darling02sur}
{Darling}, J., \& {Giovanelli}, R. 2002, \aj, 124, 100

\bibitem[{{Dickey} {et~al.}(1981){Dickey}, {Crovisier}, \& {Kazes}}]{Dickey81}
{Dickey}, J.~M., {Crovisier}, J., \& {Kazes}, I. 1981, \aap, 98, 271

\bibitem[{{Dutta} {et~al.}(2017){Dutta}, {Srianand}, {Gupta}, {Momjian},
  {Noterdaeme}, {Petitjean}, \& {Rahmani}}]{Dutta17}
{Dutta}, R., {Srianand}, R., {Gupta}, N., {et~al.} 2017, \mnras, 465, 588

\bibitem[{{Fernandez} {et~al.}(2010){Fernandez}, {Momjian}, {Salter}, \&
  {Ghosh}}]{Fernandez10}
{Fernandez}, M.~X., {Momjian}, E., {Salter}, C.~J., \& {Ghosh}, T. 2010, \aj,
  139, 2066

\bibitem[{{Fey} \& {Charlot}(2000)}]{Fey00}
{Fey}, A.~L., \& {Charlot}, P. 2000, \apjs, 128, 17

\bibitem[{{Gupta} {et~al.}(2010){Gupta}, {Srianand}, {Bowen}, {York}, \&
  {Wadadekar}}]{Gupta10}
{Gupta}, N., {Srianand}, R., {Bowen}, D.~V., {York}, D.~G., \& {Wadadekar}, Y.
  2010, \mnras, 408, 849

\bibitem[{{Gupta} {et~al.}(2013){Gupta}, {Srianand}, {Noterdaeme}, {Petitjean},
  \& {Muzahid}}]{Gupta13}
{Gupta}, N., {Srianand}, R., {Noterdaeme}, P., {Petitjean}, P., \& {Muzahid},
  S. 2013, \aap, 558, A84

\bibitem[{{Gupta} {et~al.}(2012){Gupta}, {Srianand}, {Petitjean}, {Bergeron},
  {Noterdaeme}, \& {Muzahid}}]{Gupta12}
{Gupta}, N., {Srianand}, R., {Petitjean}, P., {et~al.} 2012, \aap, 544, A21

\bibitem[{{Gupta} {et~al.}(2017{\natexlab{a}}){Gupta}, {Srianand}, {Farnes},
  {Pidopryhora}, {Vivek}, {Paragi}, {Noterdaeme}, {Oosterloo}, \&
  {Petitjean}}]{Gupta17gvlbi}
{Gupta}, N., {Srianand}, R., {Farnes}, J.~S., {et~al.} 2017{\natexlab{a}},
  ArXiv e-prints, arXiv:1712.03511

\bibitem[{{Gupta} {et~al.}(2017{\natexlab{b}}){Gupta}, {Srianand}, {Baan},
  {Baker}, {Beswick}, {Bhatnagar}, {Bhattacharya}, {Bosma}, {Carilli},
  {Cluver}, {Combes}, {Cress}, {Dutta}, {Fynbo}, {Heald}, {Hilton}, {Hussain},
  {Jarvis}, {Jozsa}, {Kamphuis}, {Kembhavi}, {Kerp}, {Kl{\"o}ckner},
  {Krogager}, {Kulkarni}, {Ledoux}, {Mahabal}, {Mauch}, {Moodley}, {Momjian},
  {Morganti}, {Noterdaeme}, {Oosterloo}, {Petitjean}, {Schr{\"o}der}, {Serra},
  {Sievers}, {Spekkens}, {V{\"a}is{\"a}nen}, {van der Hulst}, {Vivek}, {Wang},
  {Wong}, \& {Zungu}}]{Gupta17mals}
{Gupta}, N., {Srianand}, R., {Baan}, W., {et~al.} 2017{\natexlab{b}}, ArXiv
  e-prints, arXiv:1708.07371

\bibitem[{{Heiles} \& {Troland}(2003)}]{Heiles03}
{Heiles}, C., \& {Troland}, T.~H. 2003, \apj, 586, 1067

\bibitem[{{Hwang} \& {Chiou}(2004)}]{Hwang04}
{Hwang}, C.-Y., \& {Chiou}, S.-H. 2004, \apj, 600, 52

\bibitem[{{Kanekar} {et~al.}(2018){Kanekar}, {Ghosh}, \&
  {Chengalur}}]{Kanekar18}
{Kanekar}, N., {Ghosh}, T., \& {Chengalur}, J.~N. 2018, ArXiv e-prints,
  arXiv:1801.07688

\bibitem[{{Kanekar} {et~al.}(2005){Kanekar}, {Carilli}, {Langston}, {Rocha},
  {Combes}, {Subrahmanyan}, {Stocke}, {Menten}, {Briggs}, \&
  {Wiklind}}]{Kanekar05}
{Kanekar}, N., {Carilli}, C.~L., {Langston}, G.~I., {et~al.} 2005, Physical
  Review Letters, 95, 261301

\bibitem[{{Kanekar} {et~al.}(2014){Kanekar}, {Prochaska}, {Smette}, {Ellison},
  {Ryan-Weber}, {Momjian}, {Briggs}, {Lane}, {Chengalur}, {Delafosse}, {Grave},
  {Jacobsen}, \& {de Bruyn}}]{Kanekar14}
{Kanekar}, N., {Prochaska}, J.~X., {Smette}, A., {et~al.} 2014, \mnras, 438,
  2131

\bibitem[{{Kazes} {et~al.}(1989){Kazes}, {Mirabel}, \& {Combes}}]{Kazes89}
{Kazes}, I., {Mirabel}, I.~F., \& {Combes}, F. 1989, \iaucirc, 4856

\bibitem[{{Keeney} {et~al.}(2005){Keeney}, {Momjian}, {Stocke}, {Carilli}, \&
  {Tumlinson}}]{Keeney05}
{Keeney}, B.~A., {Momjian}, E., {Stocke}, J.~T., {Carilli}, C.~L., \&
  {Tumlinson}, J. 2005, \apj, 622, 267

\bibitem[{{Kollatschny} {et~al.}(1991){Kollatschny}, {Dietrich}, {Borgeest}, \&
  {Schramm}}]{Kollatschny91}
{Kollatschny}, W., {Dietrich}, M., {Borgeest}, U., \& {Schramm}, K.-J. 1991,
  \aap, 249, 57

\bibitem[{{Li} {et~al.}(2018){Li}, {Tang}, {Nguyen}, {Dawson}, {Heiles}, {Xu},
  {Pan}, {Goldsmith}, {Gibson}, {Murray}, {Robishaw}, {McClure-Griffiths},
  {Dickey}, {Pineda}, {Stanimirovi{\'c}}, {Bronfman}, {Troland}, \& {the PRIMO
  collaboration}}]{Li18}
{Li}, D., {Tang}, N., {Nguyen}, H., {et~al.} 2018, ArXiv e-prints,
  arXiv:1801.04373

\bibitem[{{Liszt} \& {Lucas}(1996)}]{Liszt96}
{Liszt}, H., \& {Lucas}, R. 1996, \aap, 314, 917

\bibitem[{{Liszt} \& {Lucas}(1999)}]{Liszt99}
{Liszt}, H., \& {Lucas}, R. 1999, in Astronomical Society of the Pacific
  Conference Series, Vol. 156, Highly Redshifted Radio Lines, ed. C.~L.
  {Carilli}, S.~J.~E. {Radford}, K.~M. {Menten}, \& G.~I. {Langston}, 188

\bibitem[{{Lockett} \& {Elitzur}(2008)}]{Lockett08}
{Lockett}, P., \& {Elitzur}, M. 2008, \apj, 677, 985

\bibitem[{{Momjian} {et~al.}(2006){Momjian}, {Romney}, {Carilli}, \&
  {Troland}}]{Momjian06}
{Momjian}, E., {Romney}, J.~D., {Carilli}, C.~L., \& {Troland}, T.~H. 2006,
  \apj, 653, 1172

\bibitem[{{Moshir} \& {et al.}(1990)}]{Moshir90}
{Moshir}, M., \& {et al.} 1990, in IRAS Faint Source Catalogue, version 2.0
  (1990)

\bibitem[{{Noterdaeme} {et~al.}(2017){Noterdaeme}, {Krogager}, {Balashev},
  {Ge}, {Gupta}, {Kr{\"u}hler}, {Ledoux}, {Murphy}, {P{\^a}ris}, {Petitjean},
  {Rahmani}, {Srianand}, \& {Ubachs}}]{Noterdaeme17}
{Noterdaeme}, P., {Krogager}, J.-K., {Balashev}, S., {et~al.} 2017, \aap, 597,
  A82

\bibitem[{{Parra} {et~al.}(2005){Parra}, {Conway}, {Elitzur}, \&
  {Pihlstr{\"o}m}}]{Parra05}
{Parra}, R., {Conway}, J.~E., {Elitzur}, M., \& {Pihlstr{\"o}m}, Y.~M. 2005,
  \aap, 443, 383

\bibitem[{{Rahmani} {et~al.}(2012){Rahmani}, {Srianand}, {Gupta}, {Petitjean},
  {Noterdaeme}, \& {V{\'a}squez}}]{Rahmani12}
{Rahmani}, H., {Srianand}, R., {Gupta}, N., {et~al.} 2012, \mnras, 425, 556

\bibitem[{{Sanders} \& {Mirabel}(1996)}]{Sanders96}
{Sanders}, D.~B., \& {Mirabel}, I.~F. 1996, \araa, 34, 749

\bibitem[{{Sharma} {et~al.}(2018){Sharma}, {Gyanchandani}, {Kulkarni}, {Gupta},
  {Pathak}, {Pande}, \& {Joshi}}]{artip18}
{Sharma}, R., {Gyanchandani}, D., {Kulkarni}, S., {et~al.} 2018, {ARTIP:
  Automated Radio Telescope Image Processing Pipeline}, Astrophysics Source
  Code Library, , , ascl:1802.004

\bibitem[{{Srianand} {et~al.}(2015){Srianand}, {Gupta}, {Momjian}, \&
  {Vivek}}]{Srianand15}
{Srianand}, R., {Gupta}, N., {Momjian}, E., \& {Vivek}, M. 2015, \mnras, 451,
  917

\bibitem[{{Srianand} {et~al.}(2012){Srianand}, {Gupta}, {Petitjean},
  {Noterdaeme}, {Ledoux}, {Salter}, \& {Saikia}}]{Srianand12dla}
{Srianand}, R., {Gupta}, N., {Petitjean}, P., {et~al.} 2012, \mnras, 421, 651

\bibitem[{{Srianand} {et~al.}(2013){Srianand}, {Gupta}, {Rahmani}, {Momjian},
  {Petitjean}, \& {Noterdaeme}}]{Srianand13dib}
{Srianand}, R., {Gupta}, N., {Rahmani}, H., {et~al.} 2013, \mnras, 428, 2198

\bibitem[{{Uzan}(2011)}]{Uzan11}
{Uzan}, J.-P. 2011, Living Reviews in Relativity, 14, 2

\bibitem[{{van Langevelde} {et~al.}(1995){van Langevelde}, {van Dishoeck},
  {Sevenster}, \& {Israel}}]{vanLangevelde95}
{van Langevelde}, H.~J., {van Dishoeck}, E.~F., {Sevenster}, M.~N., \&
  {Israel}, F.~P. 1995, \apjl, 448, L123

\bibitem[{{Wannier} {et~al.}(1993){Wannier}, {Andersson}, {Federman}, {Lewis},
  {Viala}, \& {Shaya}}]{Wannier93}
{Wannier}, P.~G., {Andersson}, B.-G., {Federman}, S.~R., {et~al.} 1993, \apj,
  407, 163

\bibitem[{{Weinreb} {et~al.}(1963){Weinreb}, {Barrett}, {Meeks}, \&
  {Henry}}]{Weinreb63}
{Weinreb}, S., {Barrett}, A.~H., {Meeks}, M.~L., \& {Henry}, J.~C. 1963, \nat,
  200, 829

\bibitem[{{Willett} {et~al.}(2011){Willett}, {Darling}, {Spoon},
  {Charmandaris}, \& {Armus}}]{Willett11}
{Willett}, K.~W., {Darling}, J., {Spoon}, H.~W.~W., {Charmandaris}, V., \&
  {Armus}, L. 2011, \apj, 730, 56

\bibitem[{{Womble} {et~al.}(1990){Womble}, {Junkkarinen}, {Cohen}, \&
  {Burbidge}}]{Womble90}
{Womble}, D.~S., {Junkkarinen}, V.~T., {Cohen}, R.~D., \& {Burbidge}, E.~M.
  1990, \aj, 100, 1785

\bibitem[{{Zwaan} {et~al.}(2015){Zwaan}, {Liske}, {P{\'e}roux}, {Murphy},
  {Bouch{\'e}}, {Curran}, \& {Biggs}}]{Zwaan15}
{Zwaan}, M.~A., {Liske}, J., {P{\'e}roux}, C., {et~al.} 2015, \mnras, 453, 1268

\end{thebibliography}
\end{document}